# EnergyScout: A Consumer Oriented Dashboard for Smart Meter Data Analytics

Nafees Ahmed and Klaus Mueller

Computer Science Department, Stony Brook University

**Abstract**

The increasing popularity of smart meters provides energy consumers in households with unprecedented opportunities for understanding and modifying their energy use. However, while a variety of solutions, both commercial and academic, have been proposed, research on effective visual analysis tools is still needed to achieve widespread adoption of smart meters. In this paper we explore an interface that seeks to balance the tradeoff between complexity and usability. We worked with real household data and in close collaboration with consumer experts of a large local utility company. Based on their continued feedback we designed EnergyScout - a dashboard with a versatile set of highly interactive visual tools with which consumers can understand the energy consumption of their household devices, discover the impact of their usage patterns, compare them with usage patterns of the past, and see via what-if analysis what effects a modification of these patterns may have, also in the context of modulated incentivized pricing, social and personal events, outside temperature, and weather. All of these are events which could explain certain usage patterns and help motivate a modification of behavior. We tested EnergyScout with various groups of people, households, and energy bill responsibilities in order to gauge the merits of this system.

## 1. Introduction

Given the increasing costs for energy and the threats of global warming, the conservation of energy has become a major concern. Considerable efforts are being made in the discovery of green energy sources, efficient energy production and transmission, and low-energy devices. At the same time, fostering consumer awareness on their personal energy use is another effective way to reduce the use of energy and drive behavior change.

For consumers, the traditional means to learn about household energy use is a single bill sent at the end of the month. Typically this bill contains a rather sparse amount of information - the total energy consumed in the current month, some reference data such as energy consumed over the last month and current year, average use per day, and the funds that are due. This information is too coarse to really afford an understanding of when and where energy was consumed, where waste has occurred, and how it could have been prevented.

Recognizing these shortcomings has led to the introduction of smart meters and other smart monitoring devices, constituting a major revolution in the energy industry. Smart meters measure the exact amount consumed during fixed time intervals (which can range from a fraction of seconds to minutes) and communicate these data to the energy utility. For the first time, households (and businesses) have the opportunity to precisely track their use of energy, relate it to their life-style, habits, inventory of appliances, and other factors, and based on this insight change their behavior to become more responsible energy consumers. The ample data streams generated from these devices have the potential to bring about the desired behavioral change. Conversely, by ways of these data, energy utilities can now precisely estimate how much energy has been, or is being used, at a certain time, on the house-hold level. This helps them in the diagnostics of power quality problems and also in setting special rates for peak and off-peak times [1], called dynamic pricing. By charging lower prices during off-peak, the utilities seek to incentivize consumers to lower their energy consumption during peak times and shift their load to off-peak. This can avoid power surges and the risk for blackouts. Due to these benefits, the deployment of smart-meters is growing at a consistent rate. As of 2016, 63 million smart meters have been installed, covering about 52% of all US households [2].

It was expected that consumers with access to such data would be responsive and eventually understand the benefits of dynamic pricing, moving away from the poor habits of the past. However, recent reports have shown that only a small fraction of household customers (6%) have opted for dynamic pricing [3]. Lack of convenient access to the stream of smart meter data has often been blamed for this. While systems have become available that use visualization as a means to provide this access, presenting such big data to average consumers has been a vast challenge. Another option is to automate appliance scheduling, possibly even by the utility. This, however, deprives the consumer of self-control and misses out on the educational benefits that the presentation of smart meter data might have.

A recent long-term study (18 months) by Schwartz et al. [4]. with seven households revealed that consumers are naturally curious about their energy consumption and so are well motivated to monitor it (see also earlier work by Darby al. [5]). Likewise, the same study also observed that users were in some sense proud of their engagement with regards to keeping track of energy use and mentioned this in their social context. This phenomenon has been exploited by the successful start-up company OPower, now acquired by Oracle, who used friendly competition among neighbors to lower energy consumption [6].

On the other hand, it was also reported [7] [8] that after a while consumers "backgrounded" the use of the displays, after they achieved better patterns of energy use. This experience is supported by more general studies which demonstrated that while digital technology can disrupt and change habitual human behavior, it remains unclear whether these changes are sustained over time [9]. These indefinite results can be valued as a challenge in designing home energy monitoring systems that are both aesthetic [10] and addictive to use, like a game [11]. Alternatively, Bartram [12] has suggested the use of ambient displays that only require a quick look to gauge momentary energy consumption but do not give detailed information. We have opted for a more data-rich visual display that seeks to be rewarding in the aesthetics and exploratory playfulness it offers.

When designing a visualization system for the masses, here the energy customers, it is important not to expect a high degree of visual literacy. Any sophisticated design, however powerful or beautiful, is poised to being not well adopted, given its steep learning curve. As was also emphasized by Darby [13], ease of understanding is a key. We painfully realized this problem ourselves when working with small focus groups and consumer experts in the development of our system, called Energy Scout. In order to find the right trade-off between functionality and usability, we ended up opting for visualization entities that are readily recognizable and reside at the lower end of the periodic table of visualization methods (PTVM) [14]. The prominence of these types of primitives was also indicated by the recent study of Borkin et al. [15]. To make up for the missing expressive power, we relied on a set of mainly low-level interactions such as filter, sort, association, and aggregate [16] and coupled them with playful interactions. This effectively moved the charts from the data visualization group into the information visualization group of the PTVM. As a final ingredient, we also exploited metaphors [17] to add intuitive semantic meanings to certain concepts. Metaphors fill a separate PTVM group.

During the development of Energy Scout we were fortunate to have access to a fairly large amount of real energy consumption data, collected over a year from a sufficiently diverse set of households. This helped us design our system within a realistic frame of reference. Also, throughout the system development we were continuously informed by the feedback of a team of consumer experts who were loosely associated with the energy utility we had contracted with.

Our paper is organized as follows. Section 2 presents background information and related work. Section 3 lists the design goals that guided our development. Section 4 describes our system and its implementation. Sections 5 and 6 present results and a discussion, respectively, and Section 7 concludes and motivates future work.

## 2. Background and Related Work

A smart meter is a digital version of the traditional electric meter. Mounted at the same location, it records energy consumption of the whole house at a particular time interval (from milliseconds to minutes) and communicates the data back to the utility companies [18]. Energy-monitors, on the other hand, are commercially sold consumer installed devices that serve the same purpose but with the potential of tracking energy at the individual device level. Some modern day smart-appliances also have such monitors pre-installed. But irrespective of the type of device used, they all track similar type of data - a time-series of energy consumption values in kWh.

### 2.1. Non-Intrusive Appliance Load Monitoring (NIALM)

Gauging the consumption of energy in households as a time-series has attracted the attention of researchers very early on. Already in the early 1990s Hart et al. [19] devised a rather sophisticated framework called Nonintrusive Appliance Load Monitoring (NIALM) which could take a measured energy profile and from it infer what appliances a house-hold used and when, as well as their individual energy consumption. Not surprisingly, the concept of NIALM has raised severe privacy concerns, even by the inventor himself [20]. And while it is difficult for smart meters to perform NIALM given the relatively long time intervals of the measurements, they can theoretically enable Nonintrusive Load Monitoring (NILM) - determining rough behavioral patterns, such as when someone is home, how many people are at home, and their activity levels in terms of energy use. Hence, the security challenges for the smart meter's Advanced Metering Infrastructure (AMI) are quite similar to those of the internet infrastructure - data transmission, data companies, and external data storage. By using it we gain convenience but we also incur some risk of privacy invasion. Therefore, just like cyber security and policy, the security of AMIs is an active topic in research and policy [21].

NIALM essentially allows what is called "disaggregated energy reports" – the attribution of consumed energy to specific devices,. such as appliances, EVs, pool pumps, and even hair dryers directly from the AMI data (without the need for device specific energy monitors with ZigBee interface or others). This can help utilities to locate inefficient devices, such as pool pumps, remotely, and it also alleviates consumers from any guesswork what device consumed how much energy and at which time. The pioneering company to delivers this kind of service is Blidgely [22] by ways of proprietary "deep pattern recognition" software. Now OPower also offers this kind of analysis as well as other companies such as Blue Line [23] and SmartEnergyGroups [24], The latter two derive the data by ways of an electric meter spin counting or other attachments.

### 2.2. Commercial Visual Interfaces for AMI Data

Smart-monitor companies like the ones listed so far and others such as WattVision [25] or eMeter (now acquired by Siemens) typically provide web or smartphone applications for consumers to see a visual representation of their energy profile. These interfaces rely heavily on real-time monitoring, direct line/bar plots of historical data and some simple aggregation, all at varied levels of design quality. Large data and software companies also entered the smart energy space and did so early on. Google PowerMeter [26] (now retired) provided a set of simple visual tools and obtain personalized recommendations to save energy. Some of these visual interfaces allow users to compare their energy consumption with historical data, either another day, week, month, or year. An interesting interface that aimed to go a step further was Microsoft Hohm [27] (also now retired). It required users to give specific information on their house and appliance usage patterns and then ran a simulation to provide energy use recommendations.

Google PowerMeter was discontinued since it did not reach the scale the company wanted to work on. But the data collected from users during the short-lived internet presence clearly indicated that simple access to such information can help consumers reduce their energy use, in this case by up to 15% [36]. Microsoft Hohm found a similar fate, also plagued by a lower-than-expected adoption rate. Unlike the PowerMeter, MS Hohm had a highly information centric interface which required significant web form-driven input from its users. This proved to be a turn-off for new users. OPower, on the other hand, serves as a vivid example of a successful, behavior stimulating information system. However, while OPower gives some hints on how energy can be saved (and the neighbor undone) users need then to anxiously await the next bill to see if their modifications actually had some positive outcome.

Our Energy Scout framework, on the other hand, enables users to see the effects of their strategies in real time, using their own data within our visual dashboard interface.

*2.3. Research Arena*

On the research end, Goodwin et al. [28] recently presented work that used energy visualization as a working example for the process of developing creative design strategies. Rodgers and Bartram [29] have worked on creating so-called ambient displays that seek to convey energy consumption in less intrusive ways. Conversely, the purpose of our system is not only to alert consumers of poor or good energy use. Rather, it is designed to aid them in understanding why, where, and when energy was consumed in their households.

Our work is also related to the visualization of timeseries data which can be challenging since temporal relationships can be highly complex. There may be delays in responses, an event may occur in spatially disjoint locations, and the direction of causal relationships is also often unclear [30]. There have been a number of graphical representations that model the temporal dynamics of data attributes as flows and streams, such as ThemeRiver [31], Stream Graphs [32], and the Stacked Area Charts of the Name Voyager app [33]. Chronological flow charts and sequential displays have also been well studied, in particular in the area of health care management and medicine [34][35]. Medical data are inherently temporal and can have multiple levels of detail. To that end, Aigner et al. [36] used illustrative abstractions to gradually transition between broad qualitative overviews of temporal data (e.g., blood pressure) to detailed, quantitative views. We also use abstraction as well as aggregation to alleviate the problems with large time windows. The Geotime system [37] places geo-temporal events along a story line into a spatial context. Finally, periodic patterns are well detected in a spiral display where interaction can help align the spiral to the recurrence of the pattern [38]. The first versions of our system also used spiral displays and it enabled us to find some interesting periodic pattern of energy use in our household data. But our user studies quickly revealed that users were confused by spiral displays. Instead we added a high degree of interaction to the more conventional displays as recommended by McLachlan et al. [39].

Finally, we also make ample use of comparative visualization. A frequent use of comparative visualization has been to in the study of parameter spaces (see e.g. [40][41]). We use it to let users compare their current energy consumption with previous (or future) time periods, and possibly with that of other households. As we strive to only use simple and consumer oriented visualization primitives, we mainly utilize shading effects to visually tag comparative structures.

Brehmer et al. [42] presented a design study for energy performance visualization of large building portfolios. Using this system, managers can switch between energy consumption or energy demand and filter or aggregate sets of buildings based on a common tag. An interesting long-term study (20 weeks) with 765 volunteer households was presented by Erickson [43]. The large number of households allowed them to rank a consumer's household within a set of 30 similar households. This rank is displayed on the consumer's dashboard along with plots of daily use and a static comparative bar chart that contrasts the current consumption with that of the previous year or the 30 similar households on a per-month basis. A questionnaire-based study showed that the interface helped the majority of users to better understand their use of energy and that they also discussed their energy use with others.

## 3. Design Process Methodology

From the onset we were interested in designing an interface that would be acceptable to a broad population of energy customers. Hence, at each step of the design process we involved either domain experts or general users to understand expectations and acceptance. In our design we employed a sequential waterfall approach which was comprised of three phases, illustrated in Figure1: (1) requirements analysis, (2) design and implementation, and (3) verification. The first two phases took place with frequent participation of a panel of consumer experts associated with a local energy utility. All of them were already deeply involved with the expansion of the smart grid, understood the need of energy customers, and had good first-hand experience with existing commercial interfaces. Finally, the last phase of our design our process was conducted with general users with varied background and needs. In the following, we summarize our design methodology and point out the differences among the three phases. The next sections will provide more detail for each.

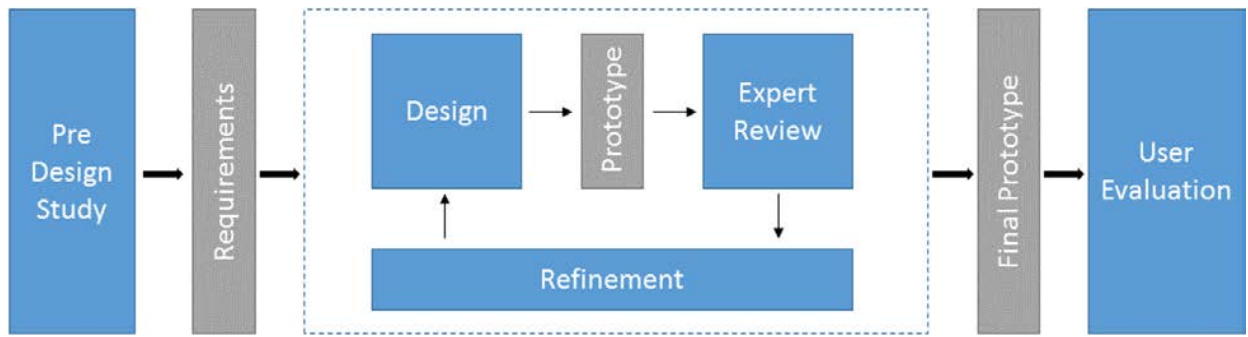

Figure 1: Overview of the design process that lead to EnergyScout.

*3.1. Phase 1 and 2: Formative Design with Experts*

The first phase formed a predesign study, essentially a requirements analysis where we sought to understand what features our system should include and how each should behave. We proceeded by analyzing the features offered by the various existing interfaces and hearing about the experiences, preferences, and desires of the members of our support panel. The following phase, phase two, interleaved the design, implementation, assessment, and refinement steps into an iterative loop. This phase was essentially a formative user study approach, routinely gathering feedback from our panel of domain experts. In each step, we would design a new element or modified an one that already existed and then we asked our panel of consumer experts to try it out. Then, depending on their feedback we kept the design as is, modified the implementation, or modified the design goal itself, and moved on to the next iterative step.

*3.2. Phase 3: Summative Evaluation with Consumers*

Once we had reached an agreement with our expert panel that the interface could be considered functioning and useful, we moved to the last phase, phase three. In this phase we performed a summative user study with human subjects of varying background and needs to evaluate the success of our overall system and design.

*3.3. Synergy of the two Design Phases*

Using the aggregate and collective knowledge of consumer experts in the design and implementation phase helped us shorten the overall design process considerably. We often engaged into a highly valuable discourse both during and after an interactive demonstration. Replaying this feedback later in the lab then helped us elucidate why a particular instance of a design was less desirable and what modifications or redesigns might work better. These new designs would then be substance for the next meeting. As such, the final phase, phase 3, constituted a validation of the overall work we accomplished with our domain experts. It not only validated our design and system, but it also validated the outcomes, effectiveness, and conclusions of the many discussions we had with our panel of experts.

## 4. Predesign Study and Requirement Analysis

The predesign study had the following goals: (1) gain an understanding of the data at hand, (2) get a sense for the mindset of energy customers, (3) determine what they hope to achieve, (4) identify a set of features that would ensure continued usage of an interface like ours.

*4.1. Source of Energy Data*

The energy consumption of household appliances can be monitored at various levels of scale. The most detailed analysis is enabled by smart-plugs, clamp-on devices, or data disaggregation which enable consumers to monitor and/or control the energy consumption of either individual devices or collections of devices. While this type of large-scale monitoring is on the rise, most AMI-equipped households (and utilities) monitor the total consumption of a household by ways of a single time-series data stream which is relatively easy to represent even in raw form. We focused our efforts to design an effective interface for these types of data. But our representations can generalize to multiple, device-centric disaggregated data streams as well.

*4.2. Getting out of the Lab: Interviews with Consumers*

We started with the assumption that people were generally concerned about their monthly energy cost and were genuinely interested in understanding their energy bill. To verify our assumption, we went to interview a set of randomly sampled people in public places in New York. In each such ad-hoc meeting, we began with an informal discussion about their energy bill. We then continued with gauging their interest in analytics interfaces and asked about any experience they had in that regard. We interviewed a total of 6 individuals, 3 male and 3 female, between 30-50 years of age. All of them were involved with the energy bills sent to their residence. We arrived at some interesting insight. We found that these consumers were left with confusions about their bills and often wondered about excessively increasing energy costs at the month's end, none was using the analysis tools that were available to them to understand the energy bill.

When asked why they did not use the tools, a general response was that they did not feel that the time required to work with the tool gave enough benefit to them. While our group was a very limited set of people, they do reflect the general mindset of the non-energy enthusiastic consumer. Even with the limited sample size, there was a clear indication that the tools were perceived as "work" and were not easy or sufficiently engaging to spark any interest in using them and gain more than the initial benefit.

*4.3. Existing Commercial Products: Market Segments*

Next, we looked into the feature sets provided by existing commercial and research solutions (see Sections 1 and 2), We identified their various features and classified them into the following five categories. In this analysis we purposely omit device-level analytics and control since our focus is on single-stream household-wide smart meter data.

**Real-time energy monitoring:** Many systems provide visualizations of spontaneous energy consumption in kWh or $, as generally displayed on the meters themselves. Some interfaces prefer to use line plots, while other interfaces make use of more aesthetic graphics and icons to make the charts more appealing to the common user. Traditional solutions makes use of dedicated display units paired with other control mechanisms. Up to date systems solutions tend to make use of mobile apps and web based interfaces to enable remote monitoring from anywhere.

Real-time monitoring in some cases also comes with provisions for alerting the user to sudden spikes or other unusual activity to make sure these won't pile up over time.

**Historical data exploration:** Most web centric solutions provide a means for exploring historical energy consumption using interactive line or bar plots. Charts for displaying the average use over some pre-defined interval are also commonly used and are preferred by users for a quick glance of their household performance.

**Energy consumption prediction:** Based on data on energy consumption, the household's energy model and other factors that might contribute such as weather and events, there are systems that provide an approximate estimate of the upcoming bill to prevent sudden surprises for the customer. However, a reasonably accurate model requires a users to be highly knowledgeable about their house and the appliances they have. Some systems provide extensive profile building (e.g. Microsoft Hohm, now discontinued), while other systems skip the cumbersome user input session and rely more on empirical data.

**Recommendations:** Some systems provide energy saving tips and recommendations to users based on their energy consumption profile as well as data from similar households. Such recommendations can be direct tips from the utility companies or they might be auto-generated based on community data generated through social interaction. Recommendation systems can have strong potential impact if implemented and promoted correctly, but they are considered more of an extended or optional feature in energy analytics.

**What-if analysis:** Some systems offer methods that allow consumers to conduct future planning, estimate approximate changes in cost/consumption that result from any modifications of the virtual house or model of behav-ior. The scale and extent of the what-if analysis may vary greatly depending on how much freedom the system allows its users in asking questions. Having a what-if analysis system available can be a great asset because it can be improved via application of visual analytics theory and techniques. A good implementation should provide a versatile and expressive analysis system, yet retain acceptable usability.

*4.4. Discussion Round with Energy Consumer Experts*

During the pre-design phase of the project, we had an open discussion session with the aforementioned group of consumer experts affiliated with a large local utility provider. The group included the company's customer technology manager, their smart-grid program lead analyst, and their customer support manager. We presented the experiences we had gained so far with regards to understanding the present solutions and customer mindset and then asked them to share their own thoughts and suggestions. The discussion that followed complemented what we had learned in the interviews with the energy consumers (Section 4.2), just now from the perspective of utility companies. First we learned that coping with increasing consumer demand and handling peak demands is a major concern for utility companies and major challenges exist with consumer awareness. *"We introduce peak/off-peak rates and other schemes to encourage our consumers change their heavy usages so that we don't have to worry about that sudden surge during peak hours of the day. We introduced smart-meters and also bought solutions for people to look at the data through our portal, but hardly anyone is using them."* Smart meters seek to promote energy awareness to consumers, helping them modify their usage behavior. But in practice this ability may serve the utility companies more than the consumers. Given the extreme cost associated with building new power plants to satisfy peak demand and the pressure associated with US government regulations requiring mindful use of energy, energy utilities have a significant interest in keeping peak demand at a reasonable level. The smart-meters and associated analytics solutions are essentially their media for communicating their mission to the average consumers, who are as of now mostly reluctant to adoption. We inquired what their thoughts were about the cause of this situation. *"My experience with the handling of customers tells me that people try out these systems once or twice, then eventually stop using them. For average consumers, the systems are not really interesting to use or may be not so intuitive. For enthusiast, they may be too basic. The system we are using, doesn't give much other information than showing you meter records from the past month. It is still better than not having any access, but I can't really tell much by just looking at those lines".* We were given access to one such system to test it ourselves. We found that even though the interface was kept simple to cater to a mass audience, even the most common operation of exploring historical data required significant clicking through drop-down date boxes. *"People are getting so much used to using those smooth and nicer apps in their phones and tablets, they find these old-school design outdated, clunky, boring and hard to use".* If we set apart the small number of energy enthusiasts who like the challenge of saving energy just for the sake of it, it became clear that making the average consumer energy-aware through these types of interfaces is a sizable challenge. The systems available so far simply require too much investment in time and effort, while offering too little reward in terms of entertainment. Possibly most amenable are usage summaries derived from the consumer's overall energy consumption gathered by smart-meters. Our challenge was to build a system along these lines, providing a sufficient degree of utility without hampering usability to build energy awareness in consumers.

**5. Prototype Design**

Informed by the findings of the pre-design study, we set about to build a prototype with the following capabilities, properties, and features:

- Focus on single-stream data as provided by the currently dominant source of household energy data - smart meters.

- Enable users to draw comparisons with historical data .

- Introduce a high degree of interaction and fluidity into the inter-face to entice users to playfully explore the data.

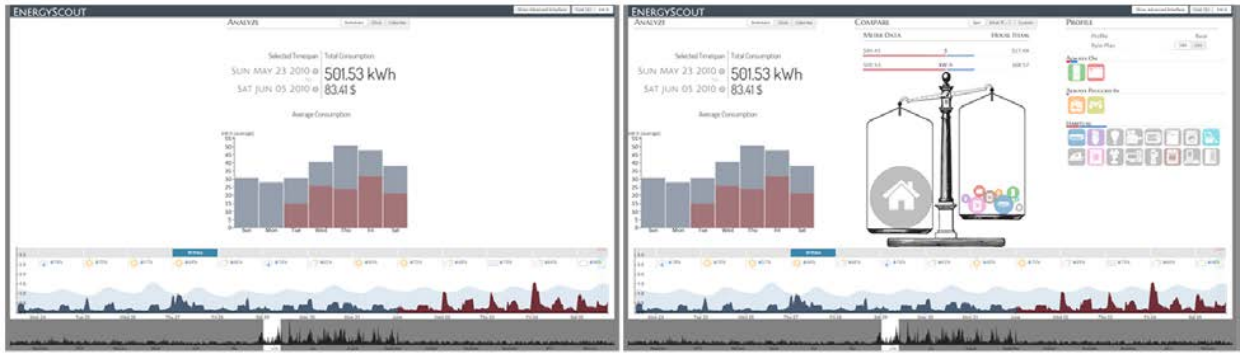

Figure 2: The Energy Scout interface showing two different perspectives: (left) basic mode for new users, and (right) advanced analysis for enthusiasts.

- Enable explanations for energy consumption by providing and linking contextual information, such as weather and calendar.

- Design an expressive what-if analysis tool that fits the need of an enthusiast but also invites the average consumer to try

- Expand the system's use to an educational tool that nudges people into thinking about energy usage and improving awareness.

We designed our interface as a dashboard organized into multiple connected visualization components, each dedicated to a particular purpose. The dashboard allows different views and perspectives (Figure 2) tailored to its audience, scalable in the dedication it requires. At the core of the dashboard are its data exploration mechanisms, which at the same time also drive other components. We experimented with various types of components to see how well they were received and then refined them accordingly for the next iteration.

### 5.1. Time-Series Exploration

Data collected from smart meters is a time series of energy consumption values. The most popular and straightforward representation of a time series is the line plot, with time mapped to the x-axis and value mapped to the y-axis. In the case at hand, it allows for an easy comprehension of the energy consumption over a particular day, hour, or even at the level of seconds. However, navigation and search within a long time-series can become a challenge. Suppose, for example, the case in which a customer receives an unusually large bill at the end of the month. The customer might not have monitored the silently streaming line plot for some time, but now he/she would pull up the analytics interface to identify the cause behind this anomaly, if it is one at all. Doing this with a traditional interface typically requires several time interval queries over the history of the data. To facilitate it, existing dashboards generally provide dropdown calendars to specify time intervals or simply provide navigation buttons to move between pages of charts. While such a data navigation approach may be suitable for one specific time interval, it becomes a tremendous bottleneck in an exploratory analysis of the type the user is facing now. The multi-level navigation interface we devised makes this (and other) data exploration tasks considerably more direct, fluid, and immersive.

#### 5.1.1. Our multi-level time series exploration interface

The interface we devised adopts an overview + detail [44] approach where we have paid great attention to the speed at which the interactions are serviced. The main component is the overview (line) chart placed at the very bottom of the interface (see Figs. 1 and 3a). It presents the entire energy consumption history of the household (within some practical limit). Users can click and drag the mouse over this overview chart to sketch out any desired time-interval or time-window (the non-grey, exposed area of the overview line chart in Figs. 2 and 3a). After selecting the time window, the corresponding data are visual-ized in a chart right above the overview chart. We did not integrate the overview and the detail chart into a single window since we wanted to preserve the detail shown in the overview window to enable quick navigation. We note that the ability to select any time interval for detail management is a capability we have not observed in existing systems (usually one can only select a fixed interval such as a day or week). It allows users to explore and compare the data over any time segment they desire, which leads to a greater degree of self-determination and empowerment [45].

Users can drag to slide/pan the time interval and use the handles on the side to adjust the size of the interval in any direction. For more precise control, mouse-wheels can also be used to zoom in and out. Such interactions can be easily mapped to equivalent touch gestures. The adjustment of the time-window can be done with finger swipes, while zooms can be mapped to a two finger pinch.

Whenever a user adjusts the time-window, every other visualization component including the detail view gets updated without much latency. As such, the time window provides a level of detail that is continuous rather than discrete. For users, this smooth and highly responsive interaction means that they can skim through the data stream within just a few seconds and figure out exactly where to focus their attention.

An important issue is the preservation of critical detail, such as minima and maxima (peaks) in the data independent of reducing zoom levels and especially in the overview plot. There is an obvious trade-off between interactive performance and accuracy that comes with down-sampling algorithms. Specifically we can do:

- Trivial down-sampling which does not preserve any extrema.

- Computing the sum of values within the down-sampled span.

- Keeping track of minima and maxima while down-sampling.

- Maintain a hierarchical representation.

The third option - keeping track of extrema - would be too time-expensive in real-time applications, and the fourth option - hierarchical decomposition - would be too wasteful in storage. Our current implementation groups the data of sub-sampled periods and generates a focus map from its sum. This turned out to be sufficiently fast, and while not fully accurate we did not come across pathological cases in which maxima fully disappeared. Eventually upon encountering a significant peak users would zoom in further and be eventually presented with the accurate values at this location.

*5.1.2. Expert panel feedback*

We presented a demo implementing the multi-level exploration interface just described to our panel of experts. They concurred that the responsiveness of the new interface could indeed provide a pathway to improve its usability and foster a more dedicated and continued use. The new interface also inspired them to propose a few new ideas. One of them was to visually differentiate between peak and consumption in both the overview and the detail plot. We implemented this by coloring the plot in red for peak hours and in blue for off-peak (see Figure 2 and Figure 3(a)).

Another comment was that we should display the data stream in actual dollar-cost, as opposed to kWh consumed, or at least provide an option for it. Other interfaces do this as well. The reason behind this being that for many consumers the unit of consumption kWh is unfamiliar or too technical, whereas the actual cost in dollars puts energy consumption into a known frame of reference, incentivizing users to lower it and spending the savings on other more desirable things.

*5.2. Integrating Contextual Information*

By exploring the historical energy data the customer will be able to locate the hour, day, or weekend that adversely contributed to the final bill. But this is only part of the solution. The ultimate goal of the smart meter is to drive behavior change. This requires isolating the cause behind the higher than normal energy use such that the customer can learn from it and prevent it from occurring in the future. Clearly, the smart meter will not know why the energy has peaked - only a device level monitoring mechanism will have a record of that. But, as dis-cussed in previous sections, device-level monitoring is still rare and so we need to bring the customer into the loop to determine the cause and activity that gave rise to the peak.

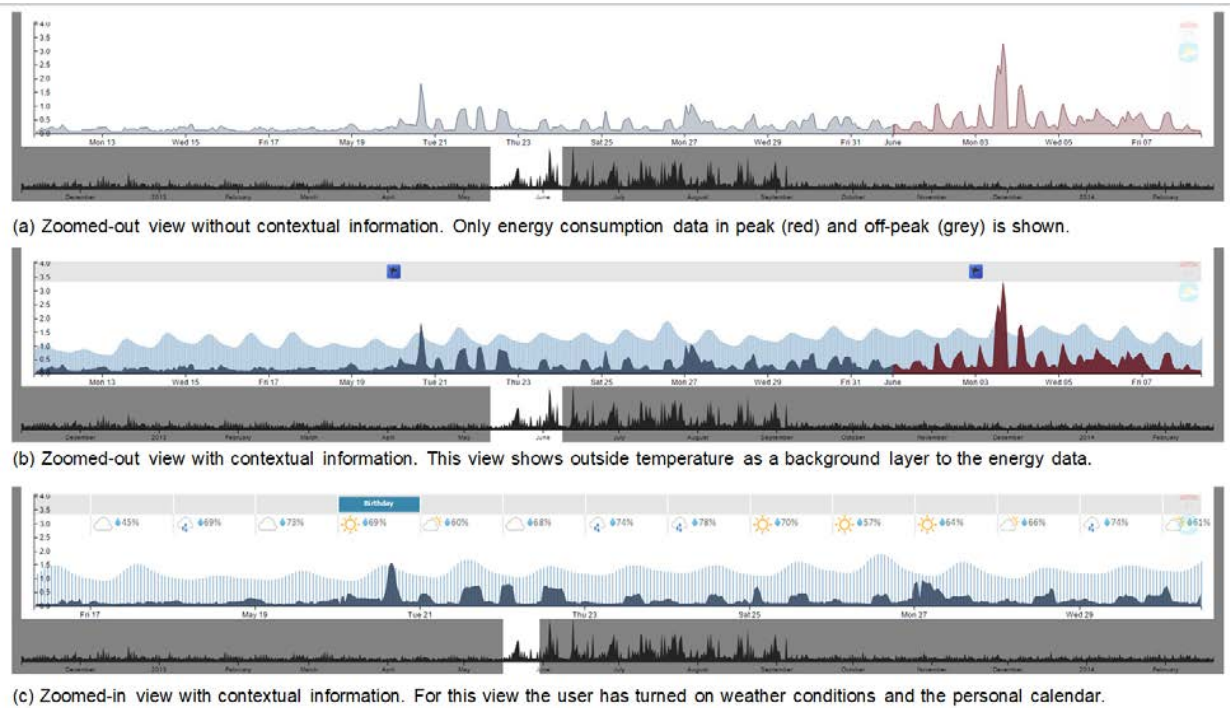

Figure 3: Data exploration with different levels of zoom and contextual data. In all figures (a)-(c), the (overview) chart on the bottom is the entire data history, whereas the (detail) chart directly above it shows the data corresponding to the time window selected by the transparent window in the overview chart. In each plot, the blue regions are due to off-peak hours, while the red regions correspond to peak hours.

Unfortunately, humans commonly do not remember everyday details, even when they are just days or a couple of weeks in the past. Let alone asking them to recall when exactly they used the water pump or when specifically they had that party months ago. They will only suspect that one or the other event may have been the cause, but this is not specific enough to trigger learning and behavior change.

Knowing the context around the house for the day under observation can greatly help investigating the cause of energy spikes. The most important context in this regard is the weather. Knowing that the day in question was the hottest day of the summer can provide a clue that the A/C might have been running the whole day. Another context source is the customer's personal calendar which may point to an early garden party which required heating the swimming pool.

*5.2.1. The Weather Context Display*

We included three types of weather parameters along with the smart meter data - temperature, humidity, and weather condition (sunny, cloudy, etc.). Similar to the energy data stream, weather information is also a time-series, now multivariate. However, the visual encoding of multiple time series in a single chart brings the potential pitfall of visual clutter [46]. Fortunately, popular consumer-end weather applications have already conformed to a certain style of visual representations for weather data which people have become familiar with. Using a similar approach brings with it the benefit of immediate acceptance and understanding. Weather applications targeted for the masses generally represent weather conditions within individual time-spans (hour/day) using well recognized icons. Temperature on the other hand is commonly represented by line/bar charts. Finally, humidity is typically communicated as a percentile with some recognizable icon to denote its type. We added a thin horizontal layer to distinguish the display from our energy data chart and display in it individual hour/day weather icons. In case of temperature, for the sake of familiarity, we used the same axis lines as the energy data. To make sure that these two information streams are well distinguishable we used three different channels of differentiation - patterns, intensity values, and color.

Figures 3(b) and 3(c) show two examples, one for a zoomed-out and one for a zoomed-in view. In these plots we can observe that high temperatures and humidity levels are somewhat related to the household's high energy consumption, but they are not the only cause. Section 5.2.4 presents a thorough investigation that is enabled by both the weather and the calendar context display discussed next

### 5.2.2. *The calendar context display*

As mentioned, another important contextual information stream is one that encodes the events related to the house. Having a calendar that holds the household events, such as holidays, parties, or any other personal recorded events provides valuable help to the customer in the quest to figure out possible causes of a consumption anomaly. However, asking a customer to maintain a separate calendar for the house can be a significant turn-off for casual users who are already not overly inclined to spend time with energy analytics systems. Hence, we synchronize our system with a user provided personal calendar. Currently, we provide support for Google calendar which is a frequently used medium for such bookkeeping and is also free. Event information from the calendar is accessible by the user in two different ways. The full calendar view mimics the look and the interactions of the web interface of Google calendar to ensure seamless transition into our system. This full calendar view is shown as an overlay to the interface on demand. However, the more appropriate encoding for aggregate analysis is to integrate this information into the energy display. For this purpose, we provide a thin layer on top of the weather bar that holds appropriate icons to denote particular events. Figure 3c shows the calendar layer on top of the display.

Note that our integration of the calendar into the energy data display is orthogonal to integrating the energy data into a calendar as was proposed by Huang et al. [47]. The latter assigns the calendar the prominent role, but this is undesirable in our case because it would make other operations on the energy data difficult.

Finally, the calendar display also doubles as a tool for annotation and quick notes. Whenever the user finds some interesting fact about home energy, he may make a quick annotation by clicking on the calendar bar. Such annotation provides more resources for future analysis and can lead to thoughtful interaction with the data [48].

### 5.2.3. *Interacting with the context displays*

In coherence with our dynamically adapting exploration interface, the detail of the weather and events chart also depends on the length of the time-interval. In our system, weather information becomes more and more detailed as one zooms in and can eventually separate into individual hours (see again Figure 3c). Events and annotations on the other hand are represented as icons. After a certain zoom factor they give detailed information about event and time-span. Finally to prevent first-time users from becoming overloaded with too much information, we provide two toggle buttons that turn on/off the weather and events overlays.

### 5.2.4. *Joe's birthday on May 20 - and its long-term effects*

The zoomed out view shown in Figure 3b indicates a set of large peaks - one peak on May 20 and a series of them in early June. The June peaks are fairly expensive since they fall into peak periods. In this context, it is also interesting to see that the peak on May 20 started off a series of moderate peaks which culminated in the large peaks in early June. However, the series of moderate peaks did not set in with a rise of overall temperature. This rise occurred 8 days earlier - on May 13 which likely coincided with the onset of summer emerging from a milder spring. Why this discrepancy, one might ask?

In search for an explanation let us zoom into this earlier period and also turn on the calendar view, shown in Figure 3c. There we ob-serve that May 20 was Joe's birthday - Joe is the father in this house-hold of 4 - and now we recall that he had invited a large group of friends on this beautiful sunny day (a memory that is supported by the plot's weather overlay). To make them all feel comfortable Joe had turned on the air conditioner - for the first time this year - plus some other devices, such as water pump for the pool, the second refrigerator for the beer, and so on. Then the following days, having experienced the comfort that the air conditioner provided, Joe and his family fell into the habit of turning it on whenever the temperature or humidity in the house got a bit high. In spring they might have run the ceiling fan, but now they energize the A/C without much thought. This is likely the main cause for the rising consumption of energy.

While the A/C is not shown in any of the charts specifically, the plots of Figure 3b and c have high potential to trigger the kind of thought process just described. By going through this exercise, Joe and his family are set to become more aware of their habits and the ramifications these have. Once recognized, behavior change (speak, rather use the ceiling fan) can then hopefully emerge.

## 5.3. Enabling Periodic and Comparative Analyses

Energy data often have inherent periodic pattern which can be due to recurrences in user behavior as well as cycles pertaining to seasons, days, weeks, and so on. Comparing these patterns over time can give a good amount of insight. We support two types of displays for this.

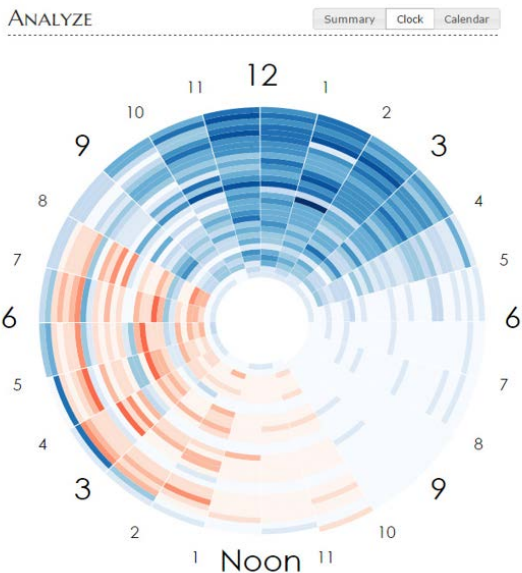

Figure 4: Displaying the smart meter data in a spiral heat map to identify periodical patterns (here, a 24 hour period). In this plot, red shaded colors encode greater energy consumption. This house-hold shows high consumption in the afternoons and evenings, some in the morning and little at night. One can also see that less energy is consumed in the two days that make up the weekend.

### 5.3.1. The spiral heat map

A powerful visualization for revealing periodic information can be obtained by wrapping the line chart into a spiral graph and use color to code the values [38]. This is confirmed by the work of Byrne et al. [49] who suggest a cyclic representation of time for infographics displays. A cyclic display subdivides a period into a meaningful set of cells, such as 24 (hour) cells for a day, 12 (month) cells or 52 (week) cells for a year, 7 (day) or 14 (half-day) cells for a week. Then it colors each cell according to the average value of the corresponding interval in the line plot. Users can choose the desired period from a menu. Figure 4 shows the spiral heat map in a 24 hour view. We have chosen a 24 hour interval (as opposed to the more familiar 12 hour display) because it better represents the re-occurring cycle of a day.

Figure 4 shows that the spiral heat map is able to reveal much interesting information not readily evident in regular line plots. The particular household that is visualized here seems to be composed of individuals that are mainly around in the afternoon and not in the weekend (see caption for more analytical detail). Other households in our collection showed clear day to day cycles, while yet others had a persistent pat-tern of staying up late on weekends and holidays. Finally there were households with seasonal effects who exhibited little energy consumption in the winter but much more in the summer. We conjectured that these households probably had oil or gas heat in the winter but used air-conditioning in the summer.

While this type of visual analysis clearly touches on privacy concerns, it can be very useful for households to discover unexplained periodic energy consumptions, such as a timer for a heater that is not set correctly. In addition, by mapping peak/off-peak rates into the spiral, it also has good potential to make poor energy use patterns visually apparent and so possibly change user behavior.

We added the spiral heat map as a new component and connected it to the exploration interface. However, when we showed the prototype to our expert panel, their response was mixed. Since radial plots and heat maps are not common in everyday use, most panel members were confused about what it might represent. When we explained its function, several members eventually liked the flexibility of providing periodic pattern discovery but yet an important objection was raised. *"People are really re-*

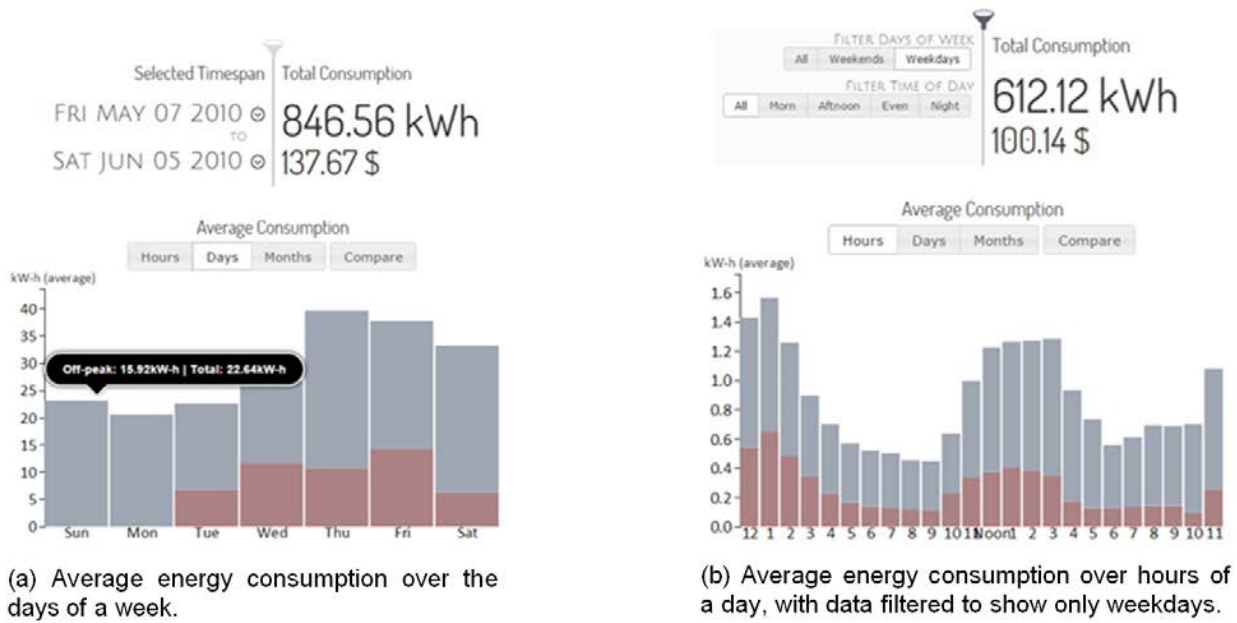

Figure 5: Summary analysis panel (scale is in kWh). The red portions are due to energy consumed during peak hours, while the blue portions are for off-peak hours. If a $-scale had been used, the red bars would have taken up a significantly larger portion.

*pulsive of something they can't quickly learn or comprehend. Having such chart in an important part of the interface may lead to rejection by the users. There may be a few who may really like it and use it, but for most it can be overwhelming"*. Consequently, we disabled this module from the common interface, but left it for use in enthusiast mode. As a side-effect this lack of understanding of the spiral heat map stimulated work on a designing a framework that would teach average users these more advanced visualization paradigms from basic ones [50].

#### 5.3.2. Interactive bar charts

During our spiral heat map presentation some members of the panel repeatedly asked *"Can't you do this with bar charts? People would like those a lot better."* This preference of bar charts in fact is quite in line with the findings of Boy et al. [51] who showed via crowd-sourced experiments that the visual literacy for bar charts is high in the general population.

Our own design experiments revealed that bar charts, when enhanced with interaction and filtering capabilities, can convey much of the information that our spiral heat maps can show. Similar to spiral heat maps, bar charts subdivide a length of time into regular discrete intervals, mapping value to height, as opposed to color. In fact, mapping value to height which is a planar variable allows for a better quantitative assessment than color. (Possibly this is an indication for the less frequent use of heat maps in general.) Similar interval and quantization selection interactions than for the spiral heat map can be defined - day view with 12 bars, week view with 7 or 14 bars, year view with 52 or 12 bars, and so on.

What bar charts cannot easily show are periodic relationships, but as mentioned, revealing these raises some concerns with regards to privacy anyhow. Instead, bar charts can average a user-defined set of periods into a single value. This obfuscates the periodicity of behavior somewhat, but still can show overall patterns. The set of selected periods for averaging can be seasonal, such as summer or winter, week or work-week, or segments of a day, such as morning, afternoon, evening, and night. Also worth mentioning is that bar charts can convey peak vs. off peak consumption better than spiral heat maps since the color retinal variable is still unused and can now be employed for the labeling of the respective portions of a bar.

Figure 5 shows two instances of the bar charts our interface can provide. In Figure 5a the user has averaged the energy consumption over a period of roughly one month in late Spring 2010 and has chosen the days view which shows the averages over each day in a week. The user learns that Sunday to Tuesday the energy consumption is relatively low and off-peak but ramps up for the later parts of the week in which also a considerable amount of expensive peak energy is used. In this figure we also see that hovering over a bar brings up some quantitative information. Likewise, Figure 5b shows an hours view where the consumption is binned into hours of a day, averaged over the same time span and all segments of a day, but only for weekdays. In this chart one can easily observe that there are two periods of high (and low) energy consumption per day - one period around 1 pm and the other around 1 am which is interesting. In fact, the energy consumption follows almost a sinusoidal pattern.

*5.3.3. Doing comparisons with the interactive bar chart*

Spiral heat maps inherently allow comparisons of time intervals at different instances of time, under the assumption that these instances are aligned with respect to the chosen period. We have implemented a similar functionality for bar charts, but generalized it beyond the capabilities of the spiral heat map. The mechanisms we provide allow comparisons of time intervals filtered by different criteria, which is not easily done with spiral heat maps. Comparative bar charts must have the same time quantization than the original bar chart. For example, if the current bar chart displays days of a week, then the comparative bar chart must also have this quantization. On the other hand, the average can be computed over a different time period or filtered by different criteria. For example, one bar chart could show the daily energy consumption average of the month just passed while the comparative chart could show the daily energy consumption average of the same month last year but averaged only over weekdays.

Comparisons are enabled by ways of pressing a "compare" button. When a comparative bar chart is activated it is initialized and combined with the current bar chart, such that for each bar the respective comparative bar is placed around it (see Figure 6. We experimented with side-by-side schemes but found that comparisons are easier with the integrative scheme we chose since it acts like a contextual background layer which it semantically is. Also, this arrangement provided us the opportunity to introduce animated transition between two states. With such transitions, users can create a natural, unsupervised mental association of markers drawn in two different states of the charts [52]. This again reinforces our design principle based on usability of the interface.

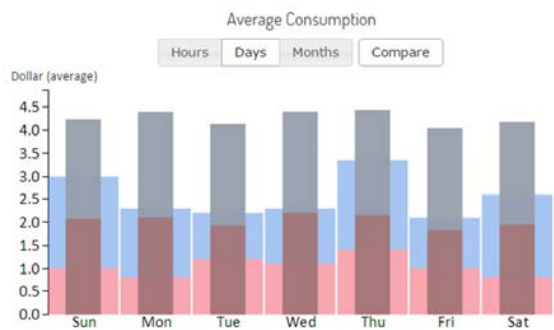

(a) Comparing the daily average of past winters (comparative chart) to the daily average of this winter (main chart).

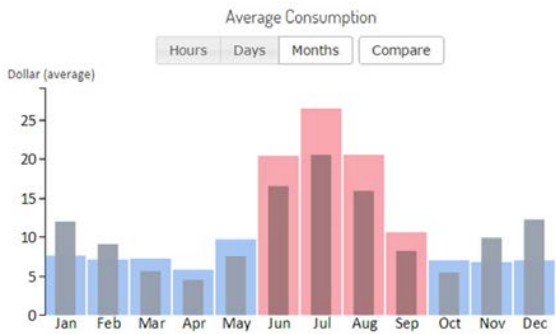

(b) Comparing the monthly average of past years (comparative chart) to the monthly average of this year (main chart).

Figure 6: Comparative bar charts. The bar chart in the foreground shows the most recent consumption while the background layer shows averages over the selected time window acting as a baseline for reference. See text for detail on the case study shown.

Figure 6(a) shows a sample comparison view where our user, Laila is taking a quick glance to see if there is anything to be concerned about. The chart she is looking at compares the daily energy consumption of the most re-cent week (during winter) to daily averages taken over similar weeks of previous winters. Clearly, the household has been consuming more energy in recent times and this requires attention. By switching to a monthly view (Figure 6(b) spanning an entire year Laila now aims to see if this has been a global trend. From this plot she learns that during the last summer their household did actually better than in the summers before - only the winter seems to the problem. This quick comparative insight gets her interested. Is it that this winter is particularly harsh compared to the past ones? Or, could it be that the heating system needs tuning, or more insulation? Now with these more detailed device-specific questions, this might be the moment at which Laila turns to the more advanced parts of the system in order to conduct further investigations and get closer to the truth. These advanced facilities are described in Section 5.4.

This case study demonstrates the propensity of the comparator to allow quick and spontaneous comparisons. One can simply hit "compare" and set up a new bar chart for comparison with the current chart. Hence, all put together, these functions bring our system quite close to the expressiveness of the spiral heat map for identifying periodic pat-terns, but without deviating too much from commonly preferred visualization techniques. And indeed, after presenting the bar chart interface to our panel, they were fully satisfied.

### 5.4. Device-Level Analytics with Non-Device Level Data

As mentioned, smart meter data in many cases do not provide the means for identifying and analyzing device-specific energy usage. Yet, the distribution of energy among household appliances represents very useful information to consumers. A possible method for understanding the device energy profiles and usage behavior in the absence of the necessary device-specific data is to engage the user in creating an approximate model of the household which can then be used to explain and predict energy household consumption. Building an energy model of the household requires two types of information: (1) details for all appliances, and (2) details on their usage. Microsoft Hohm and several other predictive model-based interfaces used form-based questionnaires to build such profiles. However, given the amount of information needed this is a rather laborious effort, and convincing hu-mans to engage in this activity has proven difficult and eventually led to the demise of these systems.

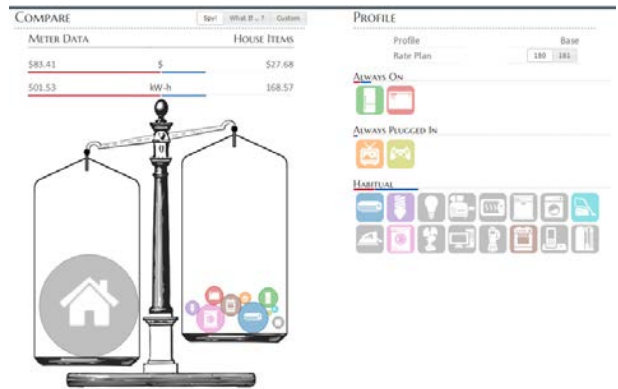

Figure 7: Turning information gathering into a game - the weight scale interface, shown here in spy mode. Users play the game by adding devices into the right bin properly scaled by duration of use and energy requirements until their cumulative weight matches the energy bill of the house, shown in the left bin. The devices located on the right hand side come from the house profile. We utilize color and icons to enable an easy association. Icons with gray colors do not have any associated events.

#### 5.4.1. Turning information gathering into an engaging game

In search for an alternative we chose an approach based on gamification. The theory of gamification promotes the idea that incentivizing users with fun and entertainment can increase their level of engagement and motivation. Encouraged by our past experiences [52][53] which used gamification for incentivizing humans to participate in visualization user studies, we aimed to use it now to add fun and entertainment into the household information gathering process. And so, similar to our gamified user study framework where we designed a game that playfully immersed the human evaluator into a visual evaluation task, we now created a game for engaging the user into the task of "filling out the household energy questionnaire".

The game we designed is based on a graphically-attractive and physics-conforming weight scale (see Figure 7) where ¢weight' is equivalent to energy consumed. Appliances are represented as circular balls or weights, whose mass and area are directly proportional to the energy consumed. As such there is a one-to-one mapping of visual and physical effect. Placing a ball onto the scale evokes a realistic simulation of the physical interactions this ball has with those already on the scale until the new configuration stabilizes. We found that by simulating the physics accurately we could greatly boost to the appeal of the scale as a game platform and so foster the level of user engagement.

Above the scale are numeric numbers coded as bar charts which display the absolute values of the weights placed on the two pans of the scale - both in kWh and dollars. The length of each bar indicates its relative value, color visualizes which side is heavier, and opacity is used to alert the user when the scale is significantly imbalanced. Placing weights onto the scale works exactly like balancing a scale in real life. In our prototype, we used the weight scale in two different modes - spy mode and what-if analysis mode. These are described next using the illustrative example of Figure 8.

*5.4.2. Spy mode - finding where the energy has gone*

In spy mode, the user is asked to think like a detective and figure out where his or her energy went. In this mode, by default there is one single weight on the scale's left pan representing the total amount of energy spent in the household. On the right of the display is the energy profiles panel with a grid of icons representing the various devices or appliances present in the household. Clicking on any of these icons (or dragging and dropping them into the scale's right pan) opens up the device profile editor. Here, users can specify the type of the appliance which at least partially hints on its usage pattern - always on (e.g. Refrigerator), always plugged in (e.g. TV) or habitual (e.g. Hair Dry-er). Users can also specify the device's approximate periodic usage and level of use by adding events. Clicking the ¢add event' opens the event editor where clock dials are available to specify the time span of usage. Toggling the days of the week on the bottom of the event editor specifies weekly periodic occurrences. Any number of events can be added which enables a good approximation of the appliance's real life usage. After the usage schedule has been set up in this manner, a device's energy consumption is computed and immediately reflected as a change in weight and size. The physics engine is then invoked to visu-ally react to these modifications.

The goal of the game is to balance the scale by putting weights (devices) onto the pan on the right until they match the weight of the household which is the big weight on the scale's left pan. We hypothesized that the challenge of balancing the scale would keep users more interested in completing the information gathering task, as opposed to the traditional questionnaire filling approach. In addition, we hypothesized that the visual grouping of devices would also foster a better understanding on how the individual devices compare in terms of their energy consumption. For example, a consumer might now suddenly realize how shockingly big the ball of an A/C unit is compared to the relatively small ball of a ceiling fan. Finally, we also split the list of devices into different categories and use a very thin bar chart along the title of the categories to represent the relative consumption among those (e.g. how much "vampire" energy is being consumed in my house compared to the total energy).

We presented the game interface to our expert panel and asked them to test it. Their initial response could be described as astonished since they were not familiar with approaches that try to turn somewhat serious work into a game. But, after spending several iterations of weight scale balancing their opinion was *"Using this weight-scale is fun for a while, but I can hardly think of a situation where I will actually seriously sit down and spend time behind perfecting my virtual house. Although, we do like how those circles make you think. Most people have no clue how much those appliances actually consume. This weight scale can be educational"*. We were encouraged by this view and subsequently left the weight scale in the interface.

*5.4.3. What-if mode - finding where energy would go*

Although the spy mode has simply the entire household on one side of the scale, our physics-driven weight scale is an effective platform for visually comparing two arbitrary sets of devices and appliances. This mode is applicable even if device-level data are available. It allows us to implement the what-if analyses or future planning sections of current dashboards in a different way than what is generally practiced. In EnergyScout, at any stage in the analysis process, the system can be switched from spy mode to what-if analysis mode. In what-if mode, the system allows comparison between two virtual households, where one of these might be the one the user built in spy mode. We call that (existing) configuration the base pro-file and its weights are kept on the right portion of the scale. Next, the user would replace the large household weight on the left portion of the scale of Figure 7 with a new profile we call what-if profile, which is initially created by simply cloning the base profile. With all the tools available in spy-mode still at our disposal, we can now take the profile in the left pan and add and remove devices, modify the usage behavior of these by modifying events, apply and try different rate plans promoted by the utility companies, and so on. Figure 8 shows an example where the user, Mark, is trying to understand if it is worth to consider changing his laundry schedule in order to save some energy. As a first interaction with the interface Mark specifies that he usually does the laundry on Friday afternoons. Then he switches to the What If interface to quickly change the laundry day to weekends and compares. As in many places with dynamic rates, Mark's utility company also considers weekends to be off-peak and the interface immediately shows the savings he can gain both in monetary and in energy units. With such an easy to use interactive scenario simulation tool, Mark now has the ability to make more in-formed decisions about his future choices. We have also experimented with a facility by which users could add provisions for exchanging the models or manufacturers of the existing devices and appliances. This can be facilitated by connecting the interface to a suitable database of products, possibly online. Endowed with such a flexibility, users can then easily ask questions like *"If I changed my old AC with that new model I heard about on TV, how much money would I save electricity each month?"* or *"People always talk about energy saving CFLs over my older bulbs, I want to see how much it really saves me if I replaced 10 of those"*. As in spy mode, any changes applied to the energy profile are immediately reflected in the weight scale. We have observed that updating the weights in the physics-based scale model gives users an engaging experience and entices them to play with these questions.

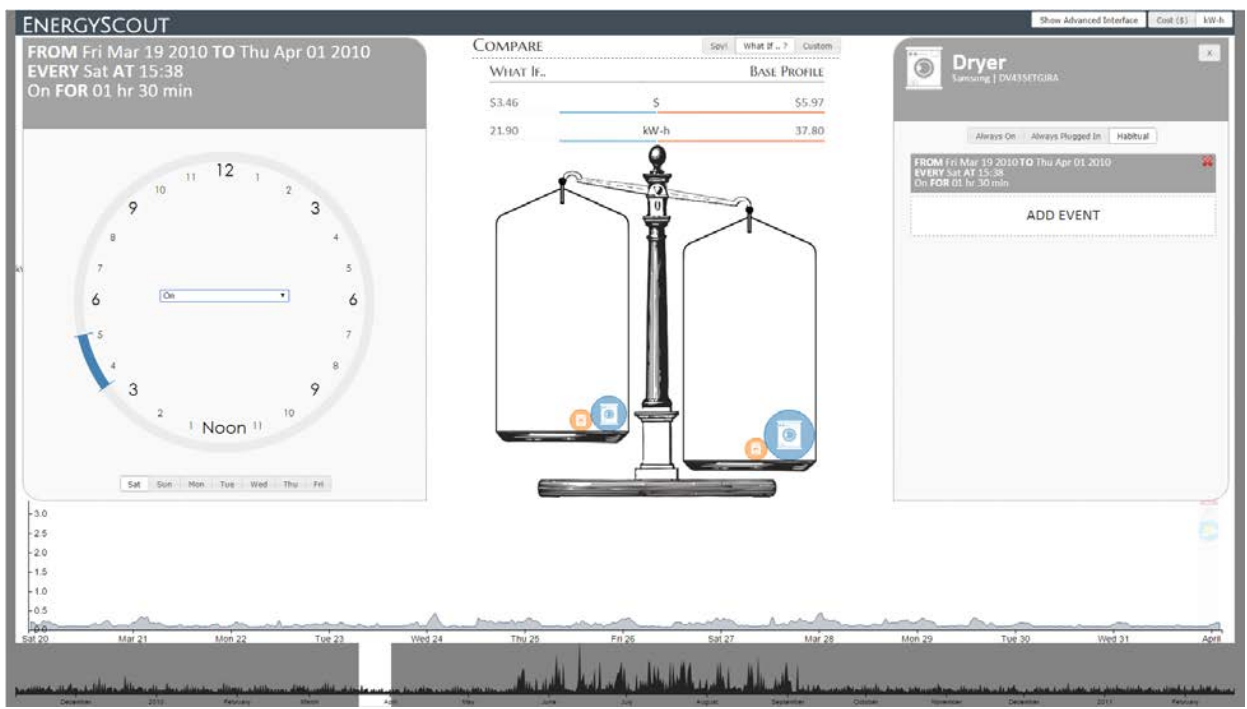

Figure 8: Future planning in EnergyScout using the what-if analysis mode. Here we conduct a comparison of two laundry scenarios to examine which of the two has greater energy efficiency and to what extent.

*5.5. Guiding New Users by Gradual Engagement*

With each prototype cycle our dashboard grew in its capabilities as new components were either freshly introduced or significantly refined. As such, being part of this evolutionary process our panellists (and defacto users) could grow their understanding of the system gradually. New users, on the other hand, would not have this luxury - they would face the entire system and all of its capabilities right away. While individually the interface components have the propensity of being intuitive in their use, having this abundance of buttons and elements together in one interface initially can be alienating to potential users to a degree where they would not use the interface at all. To overcome this problem we introduced a gradual engagement process to the user experience (UX). With this scheme, when a user first opens our dashboard, he or she is presented with only the basic part of the interface (shown in Figure 1), which consists of the exploration interface (with weather and events toggled off) and the summary bar charts without showing any buttons. Acquiring familiarity with the data representation and the interactions associated with the summary view can then be achieved with only a few swipes or clicks. In order to check out details of the summary panel touching or hovering the mouse over the chart reveals the toolbar which then can lead the user to further views and data filter options.

The gradual discovery of features and engagement has two main advantages: (1) it enables self-learning without any external tutorials, and (2) it satisfies the natural curiosity of the human mind. For many users, the analysis provided by the basic perspective of the interface is sufficient. Others, such as the energy enthusiasts, may choose to switch to the advanced analysis mode which expands the layout to expose our scale-based comparative and what-if analysis framework. Such a gradual learning scheme is analogous to the well-studied concept of flow control in game design. Controlling the flow means to continually increase a game's difficulty in manner that maintains the balance between challenge and boredom [54]. Also, as the many successful Apple products demonstrate, a well-designed consumer-facing interface should not require separate tutorials or training.

*5.6. Implementation*

Our web application is based on Java Script and HTML5 using the popular d3.js [32] library for visualization design and interactivity. To ensure a realistic physics simulation for the weight scale we used the Java Script version of the popular game library box2d [55].

One important challenge in the design of our prototype was to ensure latency-free smooth interaction and high responsiveness, even in the presence of significant amounts of data. Smart meters can produce a significant burden towards these goals when several years' worth of data needs to be processed interactively. Since the zoom-in capabilities of our system require on-demand detail up to the highest data resolution, we needed to keep all data points ready to be mapped to visual elements in the application. As web browsers have an upper limit on how many DOM elements they can process without introducing noticeable delays, we had to make sure that at any given moment the system can adapt to the platform's limitation and map visual elements at the best possible resolution. With such techniques implemented, our dashboard can provide interactive analysis at high quality even when all historical data from the first day of smart-meter installation are loaded into the system.

## 6. System Evaluation

EnergyScout provides tools geared to engage energy customers in the analysis of historical energy consumption data with the goal to help them identify patterns and anomalies based on contextual analysis. The utilization of the full potential of our system is dependent on a user's ability to associate data from his or her household with concurrent events and past experiences. An appropriate evaluation of such a dashboard would require its deployment in many households, followed by closely observing how the household members interact with the system for a longer period of time and gauging if any (positive) behavior change in energy consumption can be achieved. In our case, conducting evaluations at that scale was out of reach. Rather, we had to confine ourselves within traditional short-length user studies to verify some of our hypotheses. As mentioned, the primary goal of our prototype design phase was to create a dashboard able to provide an intuitive and engaging experience at the same or similar complexity of other consumer-grade visual analytics tools. We henceforth set out to verify if this holds true for our target users when they are exposed to the system for the very first time.

### 6.1. Our user studies with two different cohorts

We hosted two separate study sessions in different parts of the county and different social background. The first session was conducted in Long Island, NY. Among the participants were seven female and one male, with an age between 30 and 50 years. They all lived in suburban houses, under varying conditions. The second session was held in Raleigh, NC involving 9 participants - 3 female and 6 male. These individuals were of younger age and resided in rented apartments or newly bought houses.

|  | Strongly Disagree | Disagree | Neutral | Agree | Strongly Agree |
|---|---|---|---|---|---|
| easy to use | 0.00% | 0.00% | 33.33% | 45.83% | 20.83% |
| easy to learn | 4.17% | 4.17% | 20.83% | 66.67% | 4.17% |
| practical | 0.00% | 8.33% | 29.17% | 54.17% | 8.33% |

Table 1: Summary of responses from our user evaluation.

We broke up each study into two separate phases - basic exploration (Task 1) and advanced analysis (Task 2). In each phase we fol-lowed the following steps:

- A quick demonstration of the interface and its purpose.

- Let the participants discover the interface by themselves and record any comments/questions.

- Ask them to perform a specific task involving the current set of tools. Observe their interactions and record their comments/suggestions.

- Ask them to fill out a very short questionnaire.

- Open discussion among all participants for any further topic not addressed earlier.

The first three questions of the questionnaire asked about ease of use, ease of learning, and practicality of features using the Likert scale.

#### 6.1.1. Study Outcomes

Table 1 summarizes the overall responses from all 19 participants. There were almost no negative ratings, while roughly 2/3 of the test subjects expressed their general acceptance of the interface concepts. Less than 1/3 of the population was neutral. When asked which feature seemed most useful/interesting, in task 1 (basic exploration) the overall favorite was the smooth exploration scheme with mouse drag, while for task 2 (advanced analysis) the favorite was the what-if analysis. One person commented *"Zooming in out of data was fun to use. I also loved that I can get those per day information so easily"*. Some respondents identified the contextual weather and event data to be *"Very interesting and practical"*.

During the study, upon realizing that there was a way to incorporate Google calendar information, some of the subjects inquired if the system also supported other popular calendars (e.g. Microsoft Outlook, Apple Calendar), clearly showing their interest in actually associating these features. Task 2 generated keen attention from several of the participants when we showed how the weight scale can facilitate comparisons between different devices. We found that many participants, particularly females, tried to simulate many different usage behaviors and device combinations to see how they impacted cost. *"After using this software, I think it is very useful. I thought I knew about how my devices were behaving, but you can't be sure about everything without hard facts. The software cleared up some of my confusions and it was fun to learn new stuff about things that you think you have known all along."*

For some subjects though the controls available to add/modify device profiles were a turnoff. *"It was fun to juggle around with those weights, but I wouldn't have guessed how it would function if you hadn't explained it"*. We could also see diminishing interest as they struggled to fill out every little detail about energy behavior to balance the scale. To help these individuals we might add some sort of optimization scheme that does the fine-tuning of the scale. But overall, the observation that some people struggle with the scale was an expected result, already anticipated in our design phase. Trying to gather such detailed input will always be a potential pitfall for interfaces that try this. But as we found, promoting it as a playfully informative awareness building tool can elicit better acceptance rates.

We then asked the participants if they could imagine a feature that would be more useful than what we currently provide. Most suggestions centered around the refinement of font choices and colors to make them stand out even more. One participant in particular was trying to be very precise in the query of the time interval, and found the time-window to be somewhat frustrating to use for very fine adjustment cases. *"As much as I loved using the sliders, I would still prefer a simple date picker when I know the exact date/time I'm looking for"*. We agreed on this suggestion and added it as an additional feature to our interface. With regards to the device profile display and modification, some suggested that we provide an alternate list view for numeric comparisons. This is also a useful comment and in fact, all of these comments clearly show that people are generally interested in data analysis and were encouraged by our tool to engage into this activity.

We also asked about the visual design and the overall look and feel of our interface. The responses were mostly positive: *"Pleasing, easy to interpret"* and *"It is pretty user friendly. One can get enough information from this interface"*. But some of the subjects also expressed their concern about having too much content on the screen. *"The concept is good, but I think it may be too complicated for some customers to understand it without a full and clear explanation"*. As our final question, we asked the subjects if they would use this tool for their own household if we made it available. One participant said *"Absolutely"*, and seven of the 19 we worked with said *"Yes"*. But there was also a strong *"No"* from 3 participants, while the remaining subjects said *"Maybe/Sometimes"*. We believe that these are encouraging numbers which are bound to improve with further user studies and subsequent interface refinements.

## 7. Conclusions

Energy consumption data collected from smart-meters hold a valuable resource in paving the path towards a more energy efficient future. As of today, consumers are yet to adopt energy analytics at the mass level. This situation may soon change if conceptual work like our EnergyScout can lead to a better design of a finalized product. In in this paper, we explained in detail our story of the development of a consumer-facing interface, packed with exciting concepts well founded in visualization theory. The small-scale user evaluation we conducted suggests that while our design is far from being perfect, it does contain some useful techniques that were able to excite our subjects. As for future development efforts, we think it is imperative to have an extended post-deployment study with a large number of participants in order to understand the actual impact of the choices we made. Nevertheless we believe that the insights we gathered from our project as of now are already a valuable resource for future endeavors in this arena.

… References

[1] Gyamfi, S, Krumdieck, S, Urmee, T. Residential peak electricity demand responsehighlights of some behavioural issues. Renewable and Sustainable Energy Reviews 2013;25:71–77.

[2] Cooper, A. Utility-scale smart meter deployments: Building block of the evolving power grid. Inst for Electron Innovation, Washington, DC IEI Rep 2014;.

[3] Balducci, PJ, Weimar, MR, Kirkham, H. Smart grid status and metrics report. Tech. Rep.; Pacific Northwest National Lab.(PNNL), Richland, WA (United States); 2014.

[4] Schwartz, T, Stevens, G, Jakobi, T, Denef, S, Ramirez, L, Wulf, V, et al. What people do with consumption feedback: a long-term living lab study of a home energy management system. Interacting with Computers 2014;27(6):551–576.

[5] Darby, S. Making it obvious: designing feedback into energy consumption. In: Energy efficiency in household appliances and lighting. Springer; 2001, p. 685–696.

[6] Allcott, H. Social norms and energy conservation. Journal of public Economics 2011;95(9-10):1082–1095.

[7] Hargreaves, T, Nye, M, Burgess, J. Keeping energy visible? exploring how householders interact with feedback from smart energy monitors in the longer term. Energy policy 2013;52:126–134.

[8] Hargreaves, T, Nye, M, Burgess, J. Making energy visible: A qualitative field study of how householders interact with feedback from smart energy monitors. Energy policy 2010;38(10):6111–6119.

[9] Hermsen, S, Frost, J, Renes, RJ, Kerkhof, P. Using feedback through digital technology to disrupt and change habitual behavior: a critical review of current literature. Computers in Human Behavior 2016;57:61–74.

[10] Karjalainen, S. Consumer preferences for feedback on household electricity consumption. Energy and buildings 2011;43(2-3):458–467.

[11] Diakopoulos, N, Kivran-Swaine, F, Naaman, M. Playable data: characterizing the design space of game-y infographics. In: Proceedings of the SIGCHI Conference on Human Factors in Computing Systems. ACM; 2011, p. 1717–1726.

[12] Bartram, L. Design challenges and opportunities for eco-feedback in the home. IEEE Computer Graphics and Applications 2015;35(4):52–62.

[13] Darby, S. Smart metering: what potential for householder engagement? Building Research & Information 2010;38(5):442–457.

[14] Lengler, R, Eppler, MJ. Towards a periodic table of visualization methods for management. In: IASTED Proceedings of the Conference on Graphics and Visualization in Engineering (GVE 2007), Clearwater, Florida, USA. 2007;.

[15] Borkin, MA, Vo, AA, Bylinskii, Z, Isola, P, Sunkavalli, S, Oliva, A, et al. What makes a visualization memorable? IEEE Transactions on Visualization and Computer Graphics 2013;19(12):2306–2315.

[16] Pike, WA, Stasko, J, Chang, R, O'connell, TA. The science of interaction. Information Visualization 2009;8(4):263–274.

[17] Risch, JS. On the role of metaphor in information visualization. arXiv preprint arXiv:08090884 2008;.

[18] What are smart meters? http://www.whatissmartgrid.org/smart-grid-101/smart-meters; 2018. [Online; accessed 12-Feb-2018].

[19] Hart, GW. Nonintrusive appliance load monitoring. Proceedings of the IEEE 1992;80(12):1870–1891.

[20] Hart, GW. Residential energy monitoring and computerized surveillance via utility power flows. IEEE Technology and Society Magazine 1989;8(2):12–16.

[21] Yan, Y, Qian, Y, Sharif, H, Tipper, D. A survey on cyber security for smart grid communications. IEEE Communications Surveys & Tutorials 2012;.

[22] homebeat. http://homebeat.com/home; 2018. [Online; accessed 12-Feb-2018].

[23] blueline. https://www.bluelineinnovations.com; 2018. [Online; accessed 12-Feb-2018].

[24] energygroups. https://smartenergygroups.com/; 2018. [Online; accessed 12-Feb-2018].

[25] wattvision. https://www.wattvision.com/; 2018. [Online; accessed 12-Feb-2018].

[26] Power Meter. https://en.wikipedia.org/wiki/Google_PowerMeter; 2018. [Online; accessed 12-Feb-2018].

[27] Microsoft Hohm. http://en.wikipedia.org/wiki/Hohm; 2018. [Online; accessed 12-Feb-2018].

[28] Goodwin, S, Dykes, J, Jones, S, Dillingham, I, Dove, G, Duffy, A, et al. Creative user-centered visualization design for energy analysts and modelers. IEEE transactions on visualization and computer graphics 2013;19(12):2516–2525.

[29] Rodgers, J, Bartram, L. Exploring ambient and artistic visualization for residential energy use feedback. IEEE transactions on visualization and computer graphics 2011;17(12):2489–2497.

[30] Hays, DG. Isn't it about time! Counselor Education and Supervision 1977;16(3):225–228.

[31] Havre, S, Hetzler, E, Whitney, P, Nowell, L. Themeriver: Visualizing thematic changes in large document collections. IEEE transactions on visualization and computer graphics 2002;8(1):9–20.

[32] Bostock, M, Ogievetsky, V, Heer, J. $D^3$ data-driven documents. IEEE transactions on visualization and computer graphics 2011;17(12):2301–2309.

[33] voyager. http://www.babynamewizard.com/; 2018. [Online; accessed 12-Feb-2018.

[34] Plaisant, C, Milash, B, Rose, A, Widoff, S, Shneiderman, B. Lifelines: visualizing personal histories. In: Proceedings of the SIGCHI conference on Human factors in computing systems. ACM; 1996, p. 221–227.

[35] Wang, TD, Plaisant, C, Quinn, AJ, Stanchak, R, Murphy, S, Shneiderman, B. Aligning temporal data by sentinel events: discovering patterns in electronic health records. In: Proceedings of the SIGCHI conference on Human factors in computing systems. ACM; 2008, p. 457–466.

[36] Aigner, W, Miksch, S, Müller, W, Schumann, H, Tominski, C. Visual methods for analyzing time-oriented data. IEEE transactions on visualization and computer graphics 2008;14(1):47–60.

[37] Eccles, R, Kapler, T, Harper, R, Wright, W. Stories in geotime. Information Visualization 2008;7(1):3–17.

[38] Weber, M, Alexa, M, Müller, W. Visualizing time-series on spirals. In: Infovis; vol. 1. 2001, p. 7–14.

[39] McLachlan, P, Munzner, T, Koutsofios, E, North, S. Liverac: interactive visual exploration of system management time-series data. In: Proceedings of the SIGCHI Conference on Human Factors in Computing Systems. ACM; 2008, p. 1483–1492.

[40] Alabi, OS, Wu, X, Harter, JM, Phadke, M, Pinto, L, Petersen, H, et al. Comparative visualization of ensembles using ensemble surface slicing. In: Proceedings of SPIE; vol. 8294. NIH Public Access; 2012,.

[41] Malik, MM, Heinzl, C, Groeller, ME. Comparative visualization for parameter studies of dataset series. IEEE Transactions on Visualization and Computer Graphics 2010;16(5):829–840.

[42] Brehmer, M, Ng, J, Tate, K, Munzner, T. Matches, mismatches, and methods: multiple-view workflows for energy portfolio analysis. IEEE transactions on visualization and computer graphics 2016;22(1):449–458.

[43] Erickson, T, Li, M, Kim, Y, Deshpande, A, Sahu, S, Chao, T, et al. The dubuque electricity portal: evaluation of a city-scale residential electricity consumption feedback system. In: Proceedings of the SIGCHI Conference on Human Factors in Computing Systems. ACM; 2013, p. 1203–1212.

[44] Cockburn, A, Karlson, AK, Bederson, BB. A review of overview+ detail, zooming, and focus+ context interfaces. ACM Comput Surv 2008;41(1):2–1.

[45] Deci, EL, Ryan, RM. Handbook of self-determination research. University Rochester Press; 2002.

[46] Javed, W, McDonnel, B, Elmqvist, N. Graphical perception of multiple time series. IEEE transactions on visualization and computer graphics 2010;16(6):927–934.

[47] Huang, D, Tory, M, Bartram, L. Data in everyday life: Visualizing time-varying data on a calendar. Proc Poster Compendium IEEE VIS 2014;.

[48] Costanza, E, Ramchurn, SD, Jennings, NR. Understanding domestic energy consumption through interactive visualisation: a field study. In: Proceedings of the 2012 ACM Conference on Ubiquitous Computing. ACM; 2012, p. 216–225.

[49] Byrne, L, Angus, D, Wiles, J. Acquired codes of meaning in data visualization and infographics: beyond perceptual primitives. IEEE transactions on visualization and computer graphics 2016;22(1):509–518.

[50] Ruchikachorn, P, Mueller, K. Learning visualizations by analogy: Promoting visual literacy through visualization morphing. IEEE transactions on visualization and computer graphics 2015;21(9):1028–1044.

[51] Boy, J, Rensink, RA, Bertini, E, Fekete, JD. A principled way of assessing visualization literacy. IEEE transactions on visualization and computer graphics 2014;20(12):1963–1972.

[52] Archambault, D, Purchase, H, Pinaud, B. Animation, small multiples, and the effect of mental map preservation in dynamic graphs. IEEE Transactions on Visualization and Computer Graphics 2011;17(4):539–552.

[53] Ahmed, N, Zheng, Z, Mueller, K. Human computation in visualization: Using purpose driven games for robust evaluation of visualization algorithms. IEEE transactions on visualization and computer graphics 2012;18(12):2104–2113.

[54] Csikszentmihalyi, M. Flow: The psychology of optimal performance. 1990.

[55] box2d. https://github.com/erincatto/Box2D/; 2018. [Online; accessed 12-Feb-2018].